\documentstyle[11pt,newpasp,twoside,psfig]{article}
\def\edcomment#1{\iffalse\marginpar{\raggedright\sl#1\/}\else\relax\fi}
\reversemarginpar

\begin{document}
\title{UV Properties and Evolution of High-redshift Galaxies}
\author{Alberto Buzzoni}
\affil{Telescopio Nazionale Galileo, A.P.\ 565, 38700 S/Cruz de La Palma,\\
Spain, and \\
Osservatorio Astronomico di Brera, Milano, Italy}

\begin{abstract}
I assess the problem of morphological and photometric evolution of 
high-redshift galaxies in the ultraviolet wavelength range. My discussion 
will partly rely on a new set of template galaxy models, in order to 
infer the expected changes along the Hubble morphological sequence at the 
different cosmic epochs. The impact of evolution on the faint-end galaxy 
luminosity function at $z \sim 1$ and beyond will also be evaluated and 
briefly discussed.
\end{abstract}

\section{Introduction}

The exploration of galaxies at cosmological distances has sensibly changed the
current perspective of optical astronomy, as the effect of redshift is to 
allow the study of the restframe ultraviolet (UV) emission of distant objects.
The lack of a reliable overlap with local galaxy templates (mainly due to the 
atmosphere absorption), and the possible effect of evolution, require however 
a more accurate modelling of galaxy spectral energy distribution (SED) at short
wavelength in order to consistently match high-redshift observations 
(Buzzoni 2002a).

Sampling of integrated UV emission from the galaxy population at increasing 
distances has revealed to be in principle a powerful tool to track cosmic star
formation at the different cosmological epochs (Madau 1998). For its relevance,
this method has received special attention in a number of works, considering 
the possible bias sources in the data interpretation. This included the effect
of dust (Steidel et al.\ 1999; Massarotti et al.\ 2001), and the incomplete 
sampling of galaxy luminosity function at faint magnitudes 
(Connolly et al.\ 1997).
In any case, as far as evolution is taken into account, important changes are 
to be expected for galaxies beyond $z \sim 1$, both in their apparent 
morphology and UV emission. In this contribution, I would like to briefly 
assess both issues and their possible influence in the interpretation of deep 
observations.

\begin{figure}
\centerline{
\psfig{file=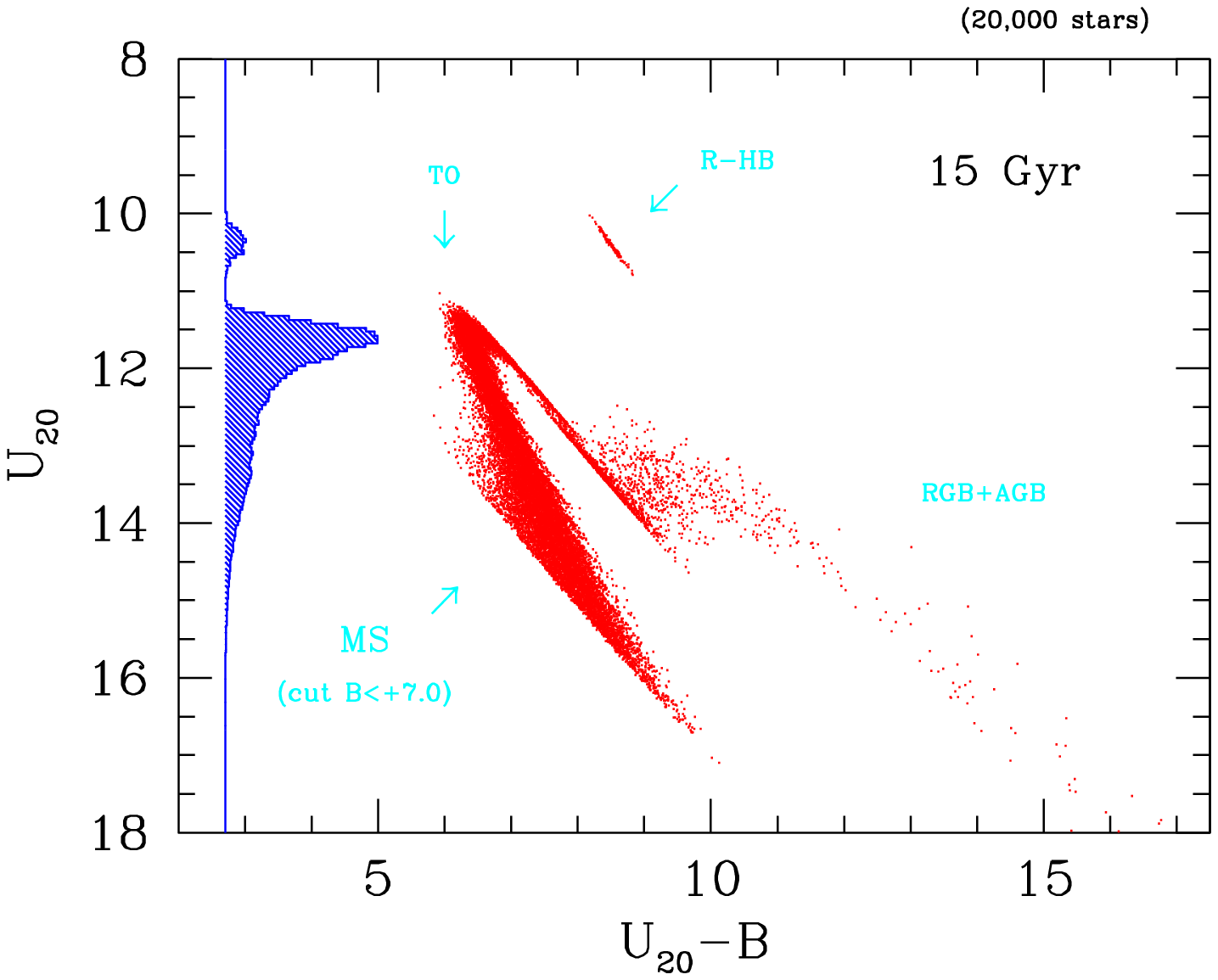,width=0.5\hsize,clip=}
\psfig{file=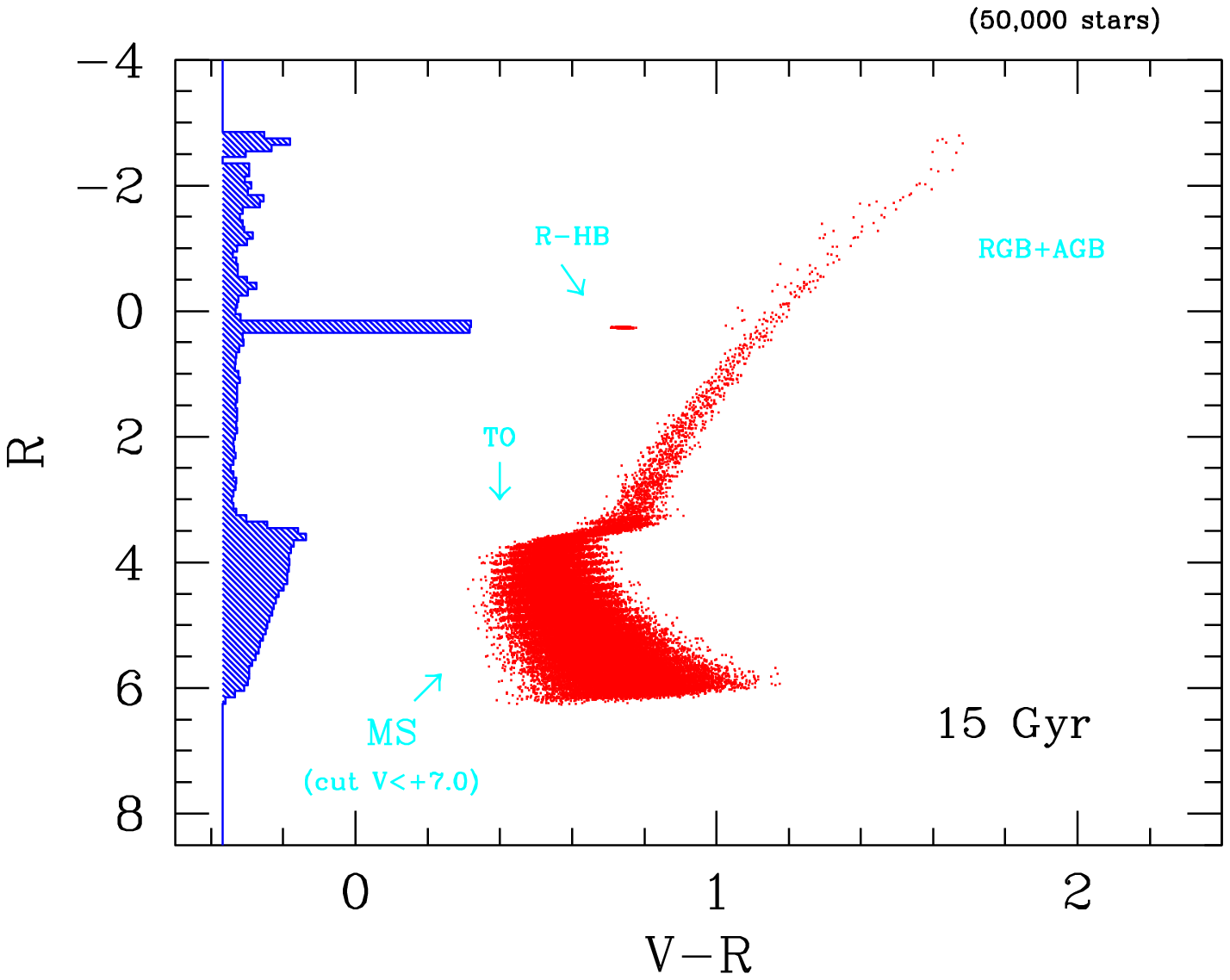,width=0.5\hsize,clip=}
}
\caption{\footnotesize{Synthetic c-m diagrams for a 15 Gyr SSP of solar 
metallicity and Salpeter IMF, after Buzzoni (1989). The left panel shows 
the magnitude distribution of stars at 2000 \AA\ ($U_{20}$), compared with 
the corresponding distribution in the Johnson $R$ band (right panel). 
The histogram on the vertical axis displays the relative fraction to the 
integrated SSP luminosity from stars in the different magnitude bins.
While in the $U_{20}$ diagram most of SSP luminosity is provided by stars 
around the turn off region, this is not the case for the $R$ diagram, where 
there is a sizable contribution from Post-MS stars (i.e.\ horizontal branch, 
R-HB, and red giant branches, RGB+AGB).}}
\end{figure}

\section{UV Emission and Star Formation Rate}

UV luminosity is known to be a fairly accurate tracer of actual star formation 
(Kennicutt 1998). This relationship basically derives from a selective 
contribution of the brightest main sequence (MS) stars to the integrated 
luminosity of a stellar aggregate. This is shown in Fig.~1, by comparing 
the theoretical c-m diagram of a simple stellar population (SSP) in two 
different photometric bands. As far as 2000 \AA\ luminosity distribution is 
concerned (left panel in the figure), one sees that a substantial fraction of 
the SSP luminosity is provided by turn off (TO) stars in the brighter 
magnitude bin of the MS region. Quite remarkably, giant stars nearly disappear 
in the ultraviolet, while on the contrary they extensively contribute to the 
integrated luminosity at optical and infrared wavelength (right panel). 
As a consequence of this selective mechanism, the integrated UV luminosity
directly relates, at any time, to the actual star formation rate, once linking
the youngest component of high-mass stars to the size of the stellar population 
as a whole, through an appropriate IMF. 
In case of a Salpeter IMF, with stars in the mass range between 0.1 and 
120~M$_\odot$, a theoretical calibration for the 2800 \AA\ luminosity 
(Buzzoni 2002b) is:
\begin{equation}
{\rm SFR}~~{\rm [M_\odot/yr]} =  {{L_{2800}}\over {4.8\,10^{27}}}
\quad {\rm [erg/sec/Hz]},
\end{equation}
quite insensitive to the details of past star formation history (see Fig.~2).

\begin{figure}
\centerline{
\psfig{file=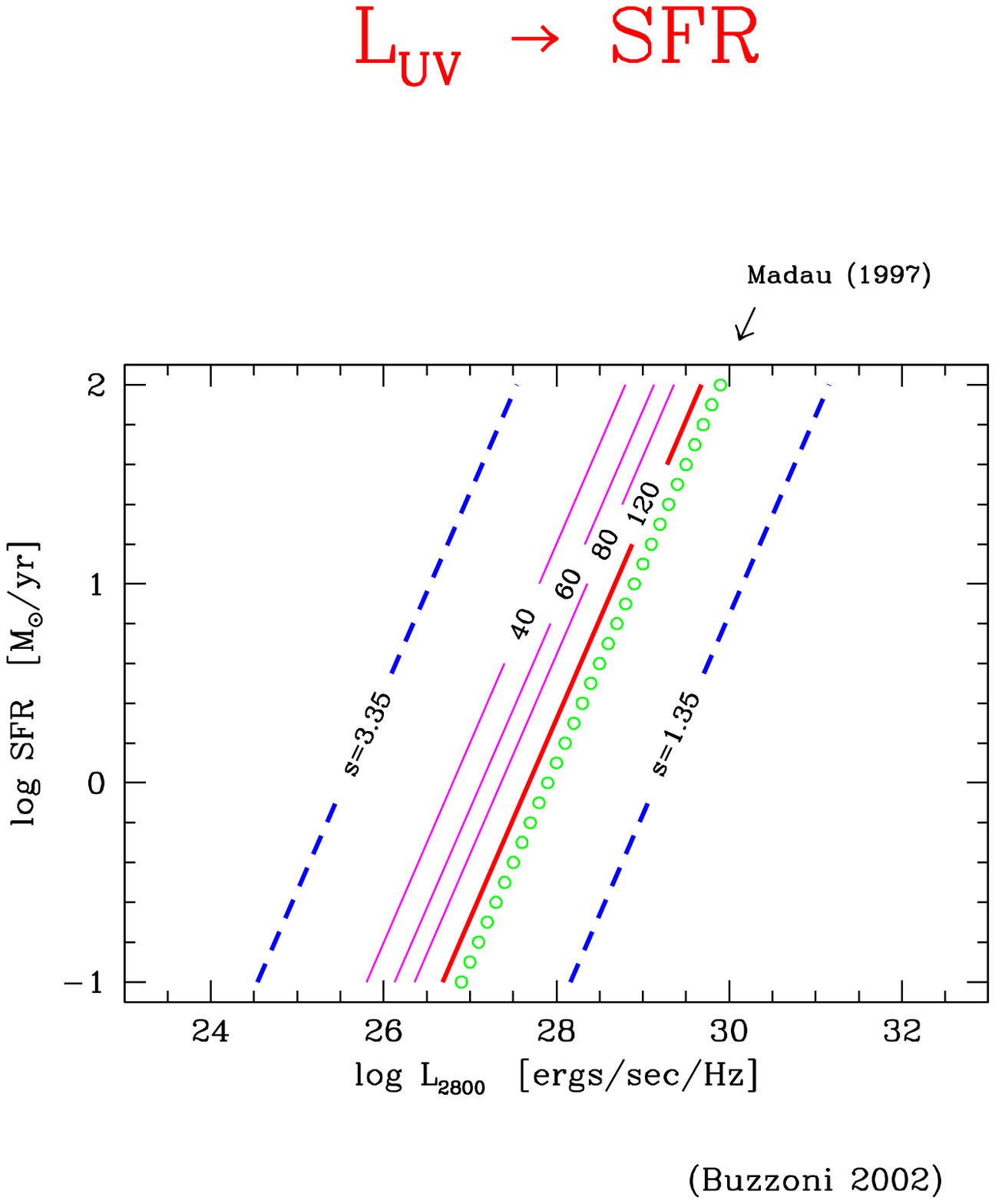,width=0.57\hsize,clip=}
}
\caption{\footnotesize{Theoretical calibration of star formation rate and 
2800 \AA\ integrated luminosity of a stellar population, after Buzzoni (2002b).
Thick solid line is the expected relationship for a Salpeter IMF with stars 
up to 120~M$_\odot$. A change of the upper cutoff mass for $M_{up} = 80, 60, 
40$~M$_\odot$ is diplayed by the thin parallel lines, as labelled. 
Dashed lines report the change in the IMF power-law slope (namely 
$dN \propto M_*^{-s}dM_*$) for dwarf- ($s = 3.35$) and giant-dominated  
($s = 1.35$) stellar populations. Dotted line is the corresponding calibration
from Madau (1997) for a Salpeter IMF ($s = 2.35$) and 
$M_{up} = 125$~M$_\odot$.}}
\end{figure}

\section {UV Morphology of High-redshift Galaxies}

Another important issue that should be considered, when tackling the study 
of the high-redshift galaxy population, is the possible morphological 
evolution. Apart from dynamical arguments (that deal, for instance, with 
the ``monolithic'' or ``hierarchical'' formation hypothesis), the apparent 
effect of redshift on the sampled SED, and the different photometric evolution 
of the galaxy sub-systems (i.e.\ bulge, disk and halo) likely tend to change 
the apparent look of galaxies with changing distance.
The are basically two ``competing'' effects, in this sense, that could 
introduce an important bias in the morphological classification with 
increasing $z$.

From one hand, in case of late-type systems, UV luminosity especially enhances
the presence of star-formation regions across the disk. When shifted to 
optical wavelength (i.e.\ for galaxies at $z \sim 1$ or beyond) this makes 
the disk plot much more knotty and irregular, compared with low-redshift 
templates (and this, {\it in spite} of any intrinsic evolution...). 
Simulations actually show that outstanding local galaxies, like M33 or M51, 
could hardly be recognized at the distance of the Hubble Deep Field 
(Burgarella et al.\ 2001; Kuchinski et al.\ 2001).
When compared with local morphological studies, the net effect of sampling 
the galaxy UV spectral range is therefore in the sense of artificially 
increasing the number of irregular (interactive?) systems at high redshift 
(van den Bergh et al.\ 2000; Kajisawa \& Yamada 2001). To some extent, 
this trend could even be reinforced by the abrupt disappearence of 
early-type systems beyond $z \sim 1.5$, mainly as a consequence of the 
disfavoring action of the k-correction.

On the other hand, we should also consider the effect of photometric evolution
when moving to large look-back times. We know that bulge stellar population 
in spirals closely resembles that of ellipticals (e.g.\ Jablonka et al.\ 1996),
and consistently matches the theoretical case of a SSP. This means that
bulge luminosity is expected to fade with time, as a consequence of the 
increasing number of dead stars and a prevailing dominance of low-mass stars.
On the contrary, the disk undergoes more or less continual stars formation
allover galaxy's life, and its luminosity is therefore dominated at any time 
by fresh (high-mass) stars.

Therefore, when looking back in time, we should in general expect a more 
prominent contribution of the bulge over the disk, and high-redshift spirals 
(and ellipticals) should more likely appear as sharply nucleated
objects, compared to their low-redshift homologues. This would explain the 
apparent lack of grand-design spirals (i.e.\ Sb-Sc systems) in the galaxy 
population at large distances (e.g.\ van den Bergh et al.\ 1996).
The effect is quantified in the two panels of Fig.\ 3, where we computed 
the relative contribution of the spheroid component (i.e.\ bulge+halo) 
to galaxy luminosity at red/infrared and ultraviolet wavelength (Johnson $U$ 
band). When accounting for expected photometric evolution, according to 
Buzzoni's (2002b) template galaxy models, one sees that later-type spirals 
(Sc-Sd types) at 1 Gyr might closely look like present-day S0-Sa systems.

\begin{figure}
\centerline{
\psfig{file=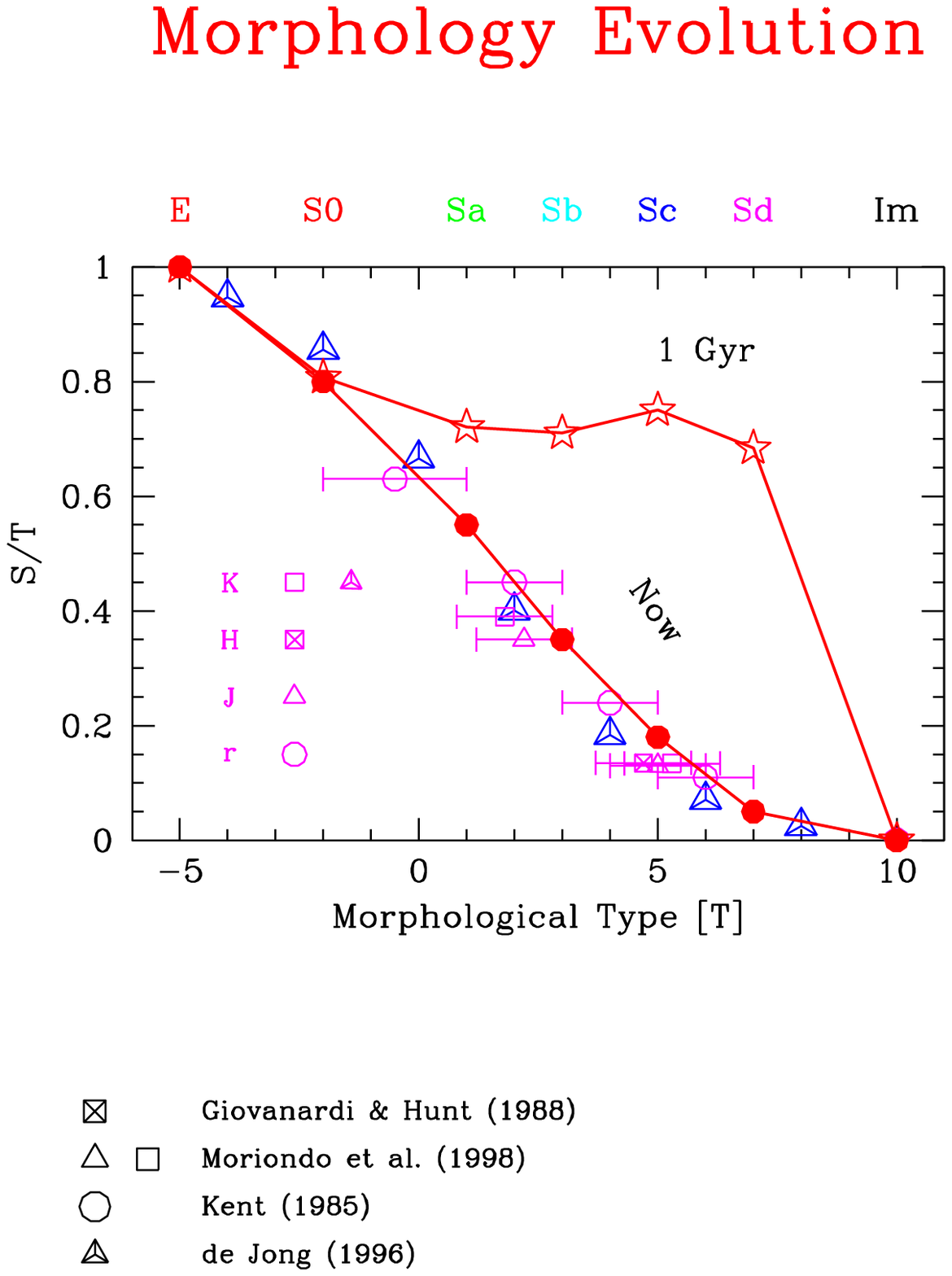,width=0.5\hsize,clip=}
\psfig{file=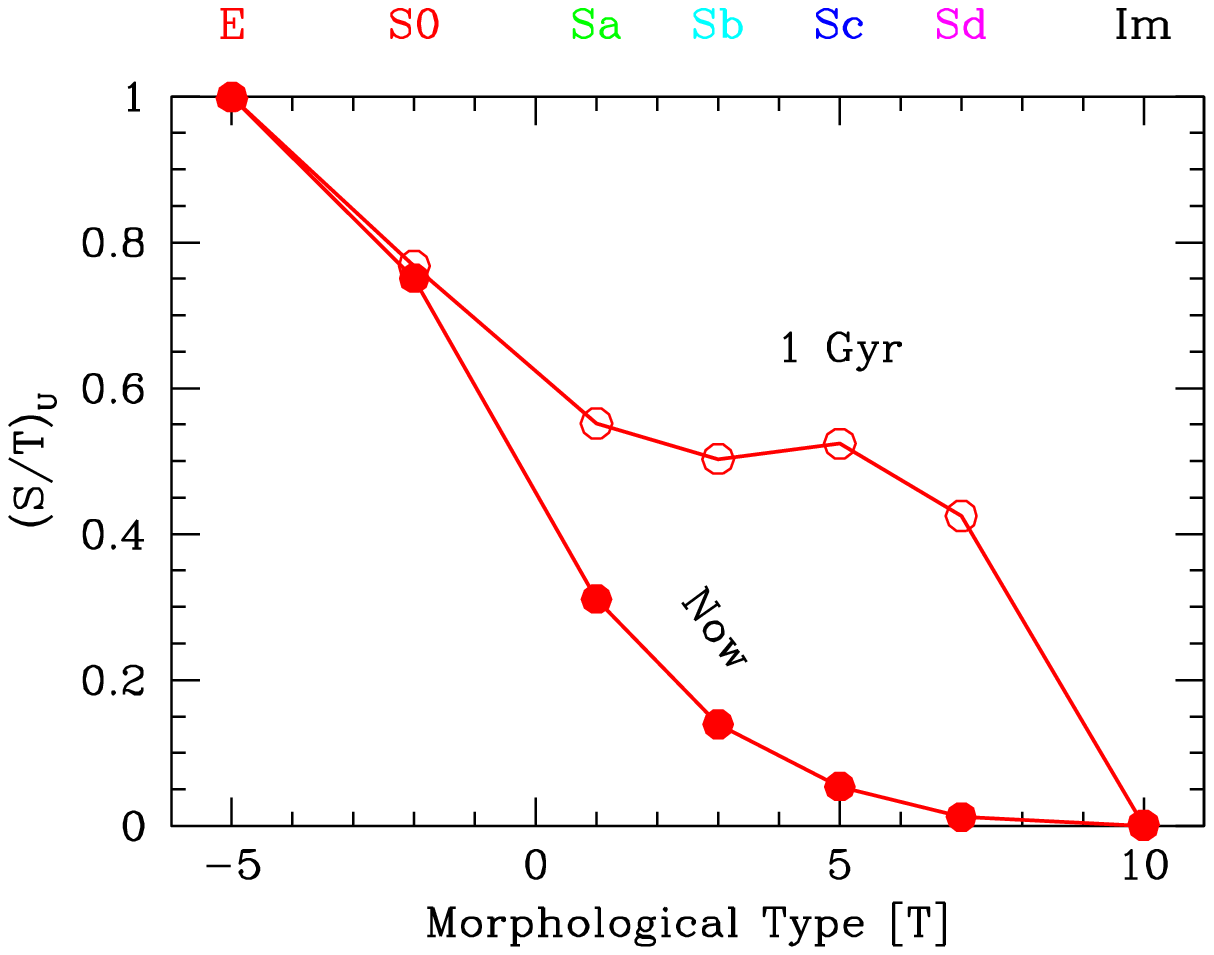,width=0.5\hsize,clip=}
}
\caption{\footnotesize{The S/T morphological parameter, defined as 
$L_{spheroid}/L_{tot}$, is studied at different ages for galaxies along 
the Hubble sequence. Left panel displays the present-day calibration (solid 
dots) at long wavelength, as derived from the data of Kent (1985; $\circ$), 
de Jong (1996; filled $\triangle$ ), Giovanardi \& Hunt (1988; filled $\sq$), 
and Moriondo et al.\ (1998; $\triangle$ and $\sq$). Its inferred trend in 
the Johnson $U$ band is plotted in the right panel, according to the template 
galaxy models of Buzzoni (2002b). The expected evolution for $t = 1$~Gyr clearly 
points to a strongly enhanced bulge contribution, especially for later-type 
spiral systems (Sc-Sd types), that would closely resemble present-day S0-Sa 
galaxies.}}
\end{figure}

\section {Expected Evolution of Galaxy Luminosity Function}

The combined action of the morphological and photometric evolution is expected
to sensibly modulate the galaxy luminosity function at high redshift.
A main issue, in this regard, concerns the shape of the faint-end tail of the
Schechter (1976) function, sizing the contribution of dwarf systems.
This problem has at least a twofold impact on the cosmological theory as 
{\it i)} faint galaxies might provide a large fraction of the UV cosmological 
background (Adelberger and Steidel 2000), thus sensibly increasing the 
estimated cosmic star formation according to the Madau (1998) plot; {\it ii)} 
low-mass systems could play a key role in the physical mechanisms that led to 
formation of the high-mass galaxies (White et al.\ 1987).

\begin{figure}
\centerline{
\psfig{file=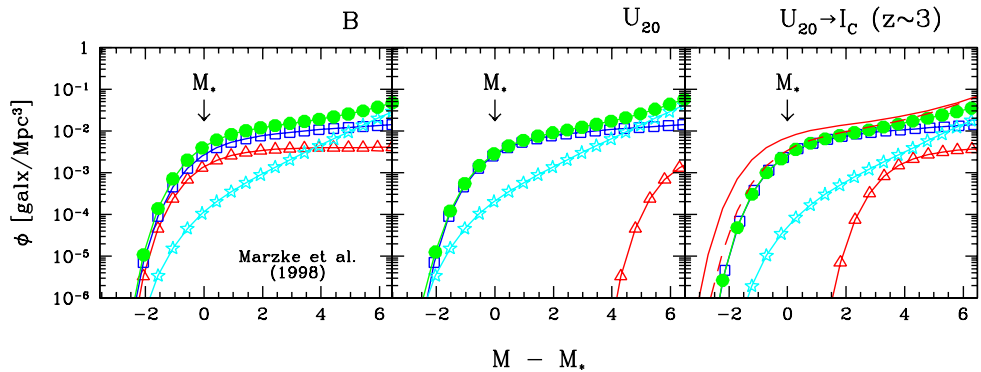,width=\hsize,clip=}
}
\caption{\footnotesize{Left panel reports the observed $B$ luminosity function 
for low-redshift galaxies, according to Marzke et al.~(1998; solid dots) and 
its partition among the different morphological types (spirals = $\sq$; 
ellipticals = $\triangle$; irregulars = $\star$).
The middle panel is the inferred galaxy distribution at 2000 \AA, according 
to the $U_{20}-B$ colors from the Buzzoni (2002b) template models, and its 
expected evolution at high redshift ($z \sim 1 \to 3$; right panel) for 5 Gyr 
galaxy models. Note that the $U_{20}$ galaxy distribution for $t =  5$~Gyr 
should roughly match the observed luminosity function at optical wavelength 
($\sim$ Johnson $B \to I$ bands). The $B$ (solid line) locus along with the 
$U_{20}$ overal galaxy distribution at $z = 0$ (dashed line) (from left and 
middle panels, respectively) are overplotted, for reference.}}
\end{figure}

As a first plain approach to the problem, we relied on the work of Marzke 
et al.\ (1998), who investigated the dependence of the local galaxy luminosity
function on morphology using a wide sample of over 5,000 low-redshift galaxies.
The Marzke et al.\ luminosity function, obtained in the Johnson $B$ band, is 
displayed in the left panel of Fig.~4, together with the relative contributions
from {\it bona fide} ``elliptical'', ``spiral'', and ``irregular'' galaxies.
The faint-end slope of the data distribution (i.e.\ the power-law index 
$\alpha$ of the Schechter function) is found to be $\alpha = -1.12$ for the 
whole sample over the observed magnitude range, but irregular galaxies clearly
dominate at fainter magnitudes with a steeper slope ($\alpha = -1.81$).
Starting from this partition, and accounting for the mean $U_{20}-B$ color as 
from the Buzzoni (2002b) template models, in the middle panel of Fig.~4 we 
report the inferred luminosity function at the 2000 \AA\ restframe wavelength.
The main feature, in this regard, is the dramatic fading of elliptical 
galaxies, as a consequence of their exceedingly ``red'' $U_{20}-B$ color 
due to a vanishing star formation at present time. Quite interestingly, this 
makes the faint-end slope of the Schechter function even steeper 
($\alpha \la -1.8$~?), as ellipticals will now reinforce the Im contribution 
at faint magnitudes.

On the other hand, if we let galaxies evolve back in time and consider 
the luminosity partition for 5 Gyr models (this is roughly the expected 
scenario for $z \sim 1 \to 3$ galaxies, depending on the cosmological model), 
then the restframe $U_{20}$ luminosity function will match observations along 
the $B \to I$ bands (cf.\ right panel of Fig.~4). Again, while the bright 
galaxy population is clearly dominated by late-type systems, and the Im galaxy
component moves to fainter magnitudes (because of a lower contribution of 
unevolved low-mass stars), ellipticals partially recover due to the age effect.
Our guess is therefore that deep optical observations should point to a 
steeper slope for the faint-end tail of the luminosity function, compared 
to the local framework (cf.\ the solid line in the right panel of Fig.~4, 
for reference), with a value of $\alpha$ eventually comprised between 
$-1.2$ and $-1.8$. 

This trend could even be strenghtened as far as the effect of dust or the 
c-m galaxy drift is taken into account, in a more sophisticated approach.
Both effects lead in fact to predict redder (i.e.\ UV fainter) galaxies at 
brighter magnitudes, and this is in the sense of steepening the faint-end 
tail of the UV galaxy luminosity function.

\acknowledgements
I am pleased to acknowledge financial support from the Italian MURST under 
grant COFIN'00 02-016.

\end{document}